\documentclass[10pt,a4paper,twocolumn,english,twocolumn, conference]{IEEEtran}
\usepackage[T1]{fontenc}
\usepackage[latin9]{inputenc}
\usepackage{float}
\usepackage{amsmath}
\usepackage{amssymb}
\usepackage{graphicx}

\makeatletter

\floatstyle{ruled}
\newfloat{algorithm}{tbp}{loa}
\providecommand{\algorithmname}{Algorithm}
\floatname{algorithm}{\protect\algorithmname}

\@ifundefined{date}{}{\date{}}
\usepackage{subfigure}
\usepackage{epstopdf}
\usepackage{cite}
\usepackage{balance}

\makeatother

\usepackage{babel}
\begin{document}

\title{Reliable Energy-Efficient Routing Algorithm for Vehicle-Assisted
Wireless Ad-Hoc Networks}

\author{\IEEEauthorblockN{Meidong Huang\IEEEauthorrefmark{1}, Bin Yang\IEEEauthorrefmark{1},
Xiaohu Ge\IEEEauthorrefmark{1}, and Wei Xiang\IEEEauthorrefmark{2}}\IEEEauthorblockA{\IEEEauthorrefmark{1}School of Electronic Information and Communications,
Huazhong University of Science and Technology, China}\IEEEauthorblockA{\IEEEauthorrefmark{2}College of Science and Engineering, James Cook
University, Australia}\IEEEauthorblockA{Corresponding author: Xiaohu Ge. Email: \IEEEauthorrefmark{1}\{meidong\_huang,yangbin,
xhge\}@hust.edu.cn, \IEEEauthorrefmark{2}wei.xiang@jcu.edu.}}
\maketitle
\begin{abstract}
We investigate the design of the optimal routing path in a moving
vehicles involved the Internet of Things (IoT). In our model, jammers
are present to interfere with the information exchange between wireless
nodes, leading to a worsened quality of service (QoS) in communications.
In addition, the transmit power of each battery-equipped node is constrained
to save energy. We propose a three-step optimal routing path algorithm
for reliable and energy-efficient communications. Moreover, results
show that with the assistance of moving vehicles, the total energy
consumed can be reduced to a large extend. We also study the impact
on the optimal routing path design and energy consumption which is
caused by the path loss, maximum transmit power constrain, QoS requirement,
etc.
\end{abstract}

\section{Introduction}

In the emerging fifth generation (5G) wireless networks, all devices
that benefit from Internet connections will be connected. Internet
of Things (IoT) technology is a key enabler of this vision by delivering
machine-to-machine (M2M) and human-to-machine communications on a
massive scale\cite{Ericsson15White}. There will be around 28 billion
connected devices by 2021, of which more than 15 billion will be M2M
and consumer-electronics devices \cite{wei2016,Ericsson15Mobility}.
The primary feature of IoT is that one device can directly link with
other devices without needing the support of infrastructure, e.g.,
base stations (BSs). Recently, increasing research efforts have been
devoted to the optimal routing design in a energy-efficient manner.

In \cite{Rahama16A}, the authors introduced a new protocol which
improves upon energy efficiency and reduces the number of dead nodes
in large-scale wireless sensor networks (WSNs). In \cite{Huynh16Delay,Tran12Minimum},
the authors proposed an algorithm to find the minimum latency and
energy-efficient path in a lossy network. Authors of \cite{Jan17A}
proposed an algorithm aiming to balance energy consumption and to
alleviate the energy hole problem. However, power constraints are
not considered in \cite{Rahama16A,Huynh16Delay,Jan17A,Tran12Minimum}
when designing the optimal routing path, which is not practical in
battery-powered networks. Additionally, in some specific scenarios
such as wireless sensors in a marine environment, BSs may not be available
to relay information. As such, these networks usually use satellites
or unmanned aerial vehicles (UAVs) to collect information. In the
future, more and more things with communications capabilities will
be mobile, e.g., the increasing number of vehicles, to assist the
in information transmission. More specifically, a vehicle can be considered
as relays to receive and forward information \cite{Bazzi13Vehicle,Korkmaz06An,Tonguz07Broadcasting,Ge16Vehicular}.
The authors of \cite{Korkmaz06An,Tonguz07Broadcasting} pay a special
attention to broadcasting in vehicular ad-hoc networks (VANET). However,
in their work, communications occur only among vehicles. While in
\cite{Bazzi13Vehicle} vehicles can communicate with the infrastructure
on the roadside in a multi-hop network.

In this paper, we investigate an ad-hoc network in suburban areas
without BSs. Nodes communicate with each other in a multi-hop way.
At the same time, there are some vehicles passing through the network
along a straight road in the network. The routing control nodes choose
the optimal path through which information is transmitted from a source
node to a destination node and determine whether to use the moving
vehicles as a mobile relay to transmit information based on the direction
of motion as well as the locations of the source node and the destination
node. This paper explores the optimal routing path design in terms
of reliability and energy efficiency in the presence of jammers \cite{Sheikholeslami14Jamming}.
Results show that the maximum power constraint and the path loss exponent
have a large impact on the routing design as well as the network performance.
The contributions of this paper are summarized as follows:
\begin{itemize}
\item We investigate the optimal routing path design in suburban areas by
jointly considering the per-node maximum transmit power constraint,
QoS, energy efficiency;
\item A three-step dynamic programming based algorithm is proposed, which
is capable of reducing total energy consumption with the assistance
of moving vehicles. 
\end{itemize}
The rest of this paper is organized as follows. Section \ref{sec:System-Model}
describes the system model, including the channel model, an analysis
of the end-to-end outage probability, and the problem formulation.
The algorithm for minimum energy consumption routing with an equal
outage probability per link based on dynamic programming is proposed
in Section \ref{sec:Algorithm}. In Section\ref{sec:Results-and-Discussions},
the simulation results are given followed by some discussions. In
the end, we conclude our paper and discuss possible future work.

\section{\label{sec:System-Model}System Model}

\subsection{\label{subsec:Network Topology}Network topology}

\begin{figure}[h]
\includegraphics[width=8.5cm]{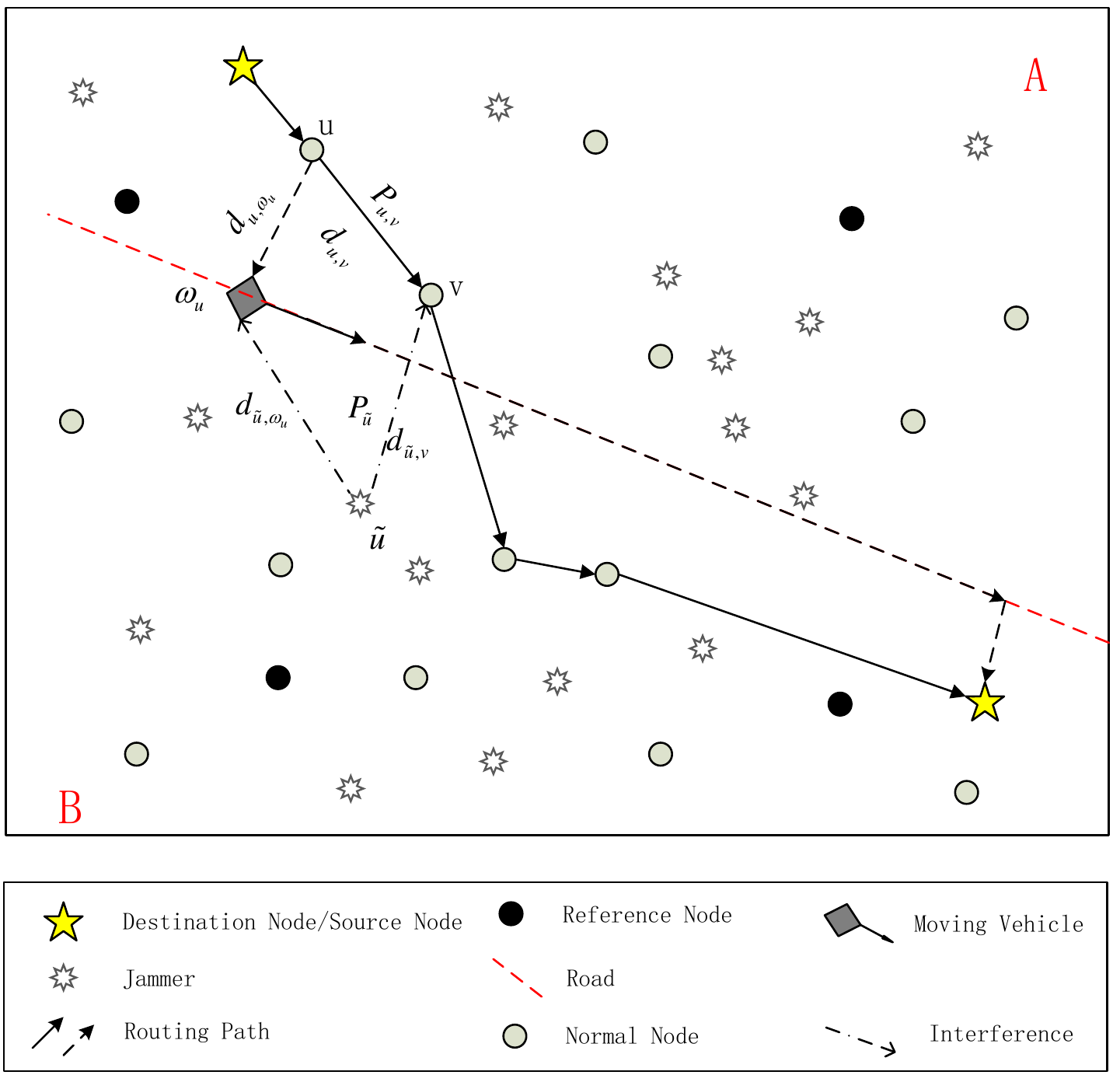}

\caption{\label{fig:System-Model}Network topology.}
\end{figure}

As illustrated in Fig.\ref{fig:System-Model}, the normal nodes depicted
in gray color exchange information among each other without GPS. However,
in order to have a good knowledge of position information of the whole
network, a few reference nodes (in black color) are equipped with
GPS \cite{Zhao16An}, which are treated as the routing control nodes
for the network. It is further assumed for the sake of simplicity
that there is only one straight road across the whole plane, on which
several vehicles are moving. Jammers which may interfere with other
nodes are randomly located in the network. It is also assumed that
each jammer is equipped with an omni-directional antenna and share
the same frequency band with the normal and reference nodes (collectively
called nodes). In this paper, reliable and low-power communications
are simultaneously considered and analyzed in consideration of the
interference of the jammers.

Assume that the locations of the nodes follow a Poisson point process
with density $\lambda_{1}$, and the locations of the jammers are
governed by another independent Poisson point process with density
$\lambda_{2}$. Denoted by $\boldsymbol{\omega}$, the location of
a moving vehicle with coordinate $\left(\omega_{x},\omega_{y}\right)$
in the plane $\mathbb{R}^{2}$, and based on the above assumptions,
the tuple $\left(\omega_{x},\omega_{y}\right)$ satisfies $a\omega_{x}+\omega_{y}+b=0$
which represents the straight road. In addition, when a moving vehicle
is transmitting (or receiving) information to (or from) normal nodes,
its location $\boldsymbol{\omega}$ is assumed to be quasi-static
as the information transmitted is of a finite size. The road divides
the plane into two parts. The source node and destination node are
on the either side of the road, respectively. We use plane A to denote
the side of the road which the source node is on, and call another
side plane B (show as Fig. 1).

Let $\Omega_{A}$ and $\Omega_{B}$ be the sets of nodes in planes
A and B respectively, of which the cardinalities are $N_{A}$ and
$N_{B}$, respectively. Let $N=N_{A}+N_{B}$. $\Omega'$ represents
the set of point on the road with a cardinality of $N'$. $\Pi$ is
the set of all possible links between two normal nodes or between
a normal node and the moving vehicle, whose cardinality is $N_{\Pi}$.
Let $\varOmega=\Omega_{A}+\Omega_{B}+\Omega'$. Then we use $G=(\Omega,\Pi)$
to denote the graph of the network. $\Im$ is the set of jammers.
Assume $u,v\in\varOmega$ and $\widetilde{u}\in\Im$ is a jammer.
Then the average outage probability from $u$ to $v$ is $P_{u,v}^{\textrm{out}}$.
Moreover, we assume that the max node transmit power is $P_{\max}$.
However, there is no power constraint for the moving vehicle. 

\subsection{\label{subsec:Problem Formulation}Problem formulation}

Frequency non-selective Rayleigh fading is assumed between any pair
of trans-receivers, including the nodes, moving vehicles and jammers.
The received signal of the link from node $u$ to node $v$ is given
as follows
\begin{equation}
{y^{(v)}}=\frac{{{h_{u,v}}\sqrt{{P_{u,v}}}}}{{d_{u,v}^{\alpha/2}}}{x^{(u)}}+\sum\limits _{\tilde{u}\in\Im}{\frac{{{h_{\tilde{u},v}}\sqrt{{P_{\tilde{u}}}}}}{{d_{\tilde{u},v}^{\alpha/2}}}{x^{(\tilde{u})}}}+n^{(v)},\label{eq:signal}
\end{equation}
where $d_{u,v}$ and $d_{\tilde{u},v}$ are the distance between nodes
$u$ and $v$ and the distance between the receiver nodes $v$ and
$\widetilde{u}$, respectively. $x^{(u)}$ and $x^{(\widetilde{u})}$
are the transmission signal from the node $u$ and jammer $\widetilde{u}$
, respectively. $P_{u,v}$ and $P_{\tilde{u}}$ are the transmit power
of $u$ and $\tilde{u}$, respectively. $h_{u,v}$ and $h_{\tilde{u},v}$
denote the channel fading from node $u$ to node $v$, and the fading
between jammer ${\tilde{u}}$ and node $v$, respectively. $\alpha$
refers to the path loss exponent, while $n^{(v)}$ indicates the noise
at receiver $v$.

Without loss of generality, we assume that $E[{\left|h_{u,v}\right|^{2}}]=1,\forall u,v\in\Omega+\Omega'$
and $E[{\left|{h_{\tilde{u},v}}\right|^{2}}]=1,\forall\tilde{u}\in\Im,v\in\Omega+\Omega'$.
In our model, because the focus of this research is on the impact
of interference on the receive signal, the noise power is ignored.
Based on the aforementioned system model, for downlink transmissions,
the SIR at the receiver node $v$ from the node u can be written by 

\begin{equation}
\textrm{SIR}_{u,v}=\frac{{{P_{u,v}}{\left|h_{u,v}\right|^{2}}d_{u,v}^{-\alpha}}}{{\sum\limits _{\tilde{u}\in\Im}{{P_{\tilde{u}}}{\left|{h_{\tilde{u},v}}\right|^{2}}d_{\tilde{u},v}^{-\alpha}}}}.\label{eq:SIR}
\end{equation}

To warrant the quality of service (QoS) of the network, the minimum
required throughput is assumed to be $\rho$. According to Shannon
theory, the threshold of the outage probability is given by

\begin{equation}
\gamma=2^{\rho}-1.\label{eq:throughput}
\end{equation}

Then outage probability with threshold $\gamma$ in our work is derived
as

\begin{align}
p_{u,v}^{{\rm {out}}} & =\Pr\left\{ \frac{{{P_{u,v}}{\left|h_{u,v}\right|^{2}}d_{u,v}^{-\alpha}}}{{\sum\limits _{\tilde{u}\in\Im}{{P_{\tilde{u}}}{\left|{h_{\tilde{u},v}}\right|^{2}}d_{\tilde{u},v}^{-\alpha}}}}<\gamma\right\} \nonumber \\
 & ={E_{{h_{\tilde{u},v}}}}\left(1-\exp\left(\frac{{-\gamma\sum\limits _{\tilde{u}\in\Im}{P_{k}}{\left|{h_{\tilde{u},v}}\right|^{2}}d_{\tilde{u},v}^{-\alpha}}}{{{P_{u,v}}d_{u,v}^{-\alpha}}}\right)\right)\nonumber \\
 & =1-\frac{1}{{\mathop\prod\limits _{k\in\Im}\left(1+\frac{{\gamma{P_{k}}d_{\tilde{u},v}^{-\alpha}}}{{{P_{u,v}}d_{u,v}^{-\alpha}}}\right)}}.\label{eq:poutj}
\end{align}

Assuming that the length of the information transmitted from $S$
to $D$ is $L$ bits, and as the transmit power and receive power
remain constant during transmission, the total consumed energy from
node $u$ to node $v$ is shown as

\begin{equation}
E_{u,v}^{\textrm{total}}=\frac{LP_{u,v}}{\rho}.\label{eq:Esum}
\end{equation}

Attributable to the independence between the hops, the outage probability
from node $S$ to $D$ is given as follows

\begin{equation}
p_{S-D}^{{\rm {\textrm{out}}}}=1-\mathop\prod\limits _{{l_{u,v}}\in{\Lambda_{S-D}}}(1-p_{u,v}^{{\rm {\textrm{out}}}}),\label{eq:PSD}
\end{equation}
where $l_{u,v}$ denotes the path from node $u$ to node $v$, $\Lambda_{S-D}$
refers to the set of paths from $S$ to $D$.

Substituting \eqref{eq:poutj} into \eqref{eq:PSD}, we arrive at
the following outage probability from $S$ to $D$ 

\begin{equation}
p_{S-D}^{\textrm{out}}=1-\mathop\prod\limits _{{\ell_{u,v}}\in{\Lambda_{S-D}}}\frac{1}{{\mathop\prod\limits _{\tilde{u}\in\Im}\left(1+\frac{{\gamma{P_{\tilde{u}}}d_{\tilde{u},v}^{-\alpha}}}{{{P_{u,v}}d_{u,v}^{-\alpha}}}\right)}}.
\end{equation}

As the nodes in the network are usually power-limited, the essential
issue is to minimize the energy consumption from $S$ to $D$, while
guaranteeing the QoS. In this context, we formulate the problem with
respect to the optimal routing path as follows
\begin{equation}
\Lambda_{\textrm{optimal}}=\underset{\Lambda\in\Lambda_{S-D}}{\arg\min}\left(E_{S-D}\left(\Lambda\right)\right),\label{eq:optimalpath}
\end{equation}
where $\Lambda_{\textrm{optimal}}$ denotes the optimal routing path
through which the energy consumption of the transmission from $S$
to $D$ is minimized, and the end-to-end outage constraint denoted
by $T$ can also be satisfied. Then we can obtain the energy consumption
$E_{S-D}$ from $S$ to $D$ as follows

\begin{align}
{E_{S-D}}(\Lambda) & =\mathop{\min}\limits _{{P_{u,v}}}\left(\sum\limits _{{l_{u,v}}\in{\Lambda_{S-D}}}{\frac{{{P_{u,v}}L}}{\rho}}\right)\nonumber \\
 & \textrm{s.t. }p_{S-D}^{\textrm{out}}\le T,0\le{P_{u,v}}\le{P_{\max}},u,v\in\Omega.
\end{align}

Then, the objective function can be derived as

\begin{align}
\Lambda_{\textrm{optimal}} & =\mathop{\arg\min}\limits _{{\Lambda_{S-D}}}\left(\sum\limits _{{l_{u,v}}\in{\Lambda_{S-D}}}{\frac{{{P_{u,v}}L}}{\rho}}\right)\nonumber \\
 & \textrm{s.t. }1-\mathop\prod\limits _{{\ell_{u,v}}\in{\Lambda_{S-D}}}\frac{1}{\prod\limits _{\tilde{u}\in\Im}\left(1+\frac{{\gamma{P_{\tilde{u}}}d_{\tilde{u},v}^{-\alpha}}}{{{P_{u,v}}d_{u,v}^{-\alpha}}}\right)}\le T,\nonumber \\
 & \qquad u,v\in\Omega,0\le{P_{u,v}}\le{P_{\max}}.\label{eq:objective function}
\end{align}

Similar to the situation in which we need to find the routing path
when the end-to-end delay is bounded \cite{Wang96Quality}, the problem
in this paper cannot be solved by traditional shortest path algorithms
such as the Dijkstra and Bell-Ford algorithms. There are some ways
to tackle this problem. The first one is to enumerate all possible
solutions and then to identify the best routing path that minimizes
energy consumption. However, in this problem, the transmission power
is continuous. That is so-called NP-complete problem. So, we cannot
find the best solution in this way. Secondly, the authors in \cite{Sheikholeslami14Jamming}
proposed an algorithm termed the Minimum Energy Routing With Approximate
Outage Per Link (MER-AP) algorithm, which applies the Lagrange multipliers
technique to assign each link power a certain expression formula.
But in this paper , the transmission power is bounded, while the transmission
power in \cite{Sheikholeslami14Jamming} is a function of the distance
of each link, the path loss exponent as well as the interference of
jammers, which may surpass the constraint of the max transmission
power. As a result, MER-AP is not suitable for this paper\textquoteright s
problem. The last one is to obtain an approximate expression and use
the Dijkstra algorithm or other methods to derive a sub-optimal solution.

\section{\label{sec:Algorithm}Optimal Routing Path Algorithm}

In this section, we propose a three-step algorithm to find the optimal
routing path such that the total energy consumption is minimized,
while guaranteeing the end-to-end outage constraint. Before detailing
our proposed algorithm, some related assumptions should be addressed
first.

\textbf{Assumption 1:} In this paper, we assume the total energy consumption
of the network does not include the vehicle\textquoteright s energy
consumption. This is because the moving vehicle is not considered
as part of the network, so its energy consumption will not be taken
into account in the objective function.

\textbf{Assumption 2:} The vehicle just communicates with its closest
node in plane B.

Assuming the fixed node $u$ communicates with the moving vehicle,
the average outage probability can be obtained as follows

\begin{equation}
p_{u,\omega_{u}}^{\textrm{out}}=1-\frac{1}{{\mathop\prod\limits _{\tilde{u}\in\Im}\left(1+\frac{{\gamma{P_{\tilde{u}}}d_{\tilde{u},v}^{-\alpha}}}{{{P_{u,{\omega_{u}}}}d_{u,{\omega_{u}}}^{-\alpha}}}\right)}},
\end{equation}
where $\omega_{u}$ is the point where the fixed node $u$ communicates
with the moving vehicle, and $\omega_{u}\in\varOmega'$. If proper
routing is ensured, the moving vehicle can act as a relay in the network
to transmit information. In addition, the moving vehicle can also
carry information over a long distance before transmitting it to the
fixed nodes in plane B. The total energy can be saved to a great extent.
However, to meet the end-to-end outage constraint as well as making
our considered scenario more practical, the locations where the vehicle
receives information from the fixed node $u$ in plane A should be
selected wisely, which should satisfy the following

\begin{equation}
(\omega_{x}',\omega_{y}')=\arg\max\limits _{\omega_{u}}\left(p_{u,\omega_{u}}^{\textrm{out}}\right).
\end{equation}

As the energy consumed by the moving vehicle is not considered, the
optimal routing path is actually divided into two sub-paths, i.e.,
from $S$ to the moving vehicle and from the moving vehicle to $D$.
Intuitively, the two sub-paths can be obtained in two separate planes,
i.e., planes A and B, as illustrated in Fig. 1. To reduce the complexity
of identifying the optimal routing path, we assume that each hop along
the routing path has an equal outage constraint, i.e.,

\begin{equation}
p_{u,v}^{{\rm {\textrm{out}}}}(m)=1-\sqrt[m]{{1-T}},st.{l_{u,v}}\in{\Lambda_{S-D}},\label{eq:equal pout}
\end{equation}
where $m$ is the number of hops. As can be seen from (\ref{eq:equal pout}),
the transmit power of each hop $p_{u,v}^{{\rm {out}}}(m)$ is highly
related to the number of hops, which is unknown in our model. Conditioned
on $m$, the optimal sub-path in plane A which is denoted as $\Lambda_{optimal}^{A}(n)$
with a $n$-hop (n=1,2...,m-1) path, and the optimal sub-path in plane
B denoted as $\Lambda_{optimal}^{B}(m-n)$ can be found using our
proposed algorithm, in which the number of hops in plane B is $m-n$.
After searching all possible $m$, the optimal routing path then is
attainable. Based on the above analysis, we propose a three-step dynamic
programming based algorithm to find the optimal routing path.

\subsection{\label{subsec:Signal-Propagation-Model-3}Routing Algorithm in Plane
A}

\begin{algorithm}[h]
\caption{Dynamic Programming Routing Selection on Plane A}

\ 1:\textbf{ for} \textbf{all} $u,v\in{\Omega_{A}},{P_{S,u}}\leqslant{P_{\max}}$\textbf{do} 

\ 2: \ \ ${C_{S-u}}(1)\left|{_{{\Pi_{S-u}}}}\right.={P_{S,u}}\cdot L/\rho$

\ 3: \textbf{end for} 

\ 4:\textbf{ for} all $u,v,u'\in{\Omega_{A}},{P_{u,v}}\leqslant{P_{\max}}$
\textbf{do }

\ 5: \ \ \textbf{for} i=2 to n-1 \textbf{do}

\ 6: \ \ \ \ $u'=\mathop{\arg\min}\limits _{u}({C_{S-u}}(i-1)+{P_{u,v}}\cdot L/\rho)$

\ 7: \ \ \ \ ${C_{S-v}}(i)={C_{S-u'}}(i-1)+{P_{u',v}}\cdot L/\rho$

\ 8: \ \ \ \ ${\varPi_{S-v}}(i)$=${\varPi_{S-u'}}(i-1)+l_{u',v}$

\ 9: \ \ \textbf{end for }

10: \textbf{end for }

11:\textbf{ for} \textbf{all} $u,\in{\Omega_{A}},{P_{u,{\omega'_{u}}}}\leqslant{P_{\max}}$\textbf{do} 

12:\ \ ${C_{S-{\omega'_{u}}}}(n)={C_{S-u}}(n-1)+{P_{u,{\omega_{u}}}^{\min}}\cdot L/\rho$

13: \ ${\varPi_{S-{\omega'_{u}}}}(n)$=${\varPi_{S-u}}(n-1)+l_{u,{\omega'_{u}}}$

14:\textbf{ end for }

15:\textbf{ return} $\Lambda_{optimal}^{A}(n)=\mathop{\arg\min}\limits _{{\Pi_{S-{\omega'_{u}}}}(n)}({C_{S-{\omega'_{u}}}}(n))$=${\varPi_{S-{\omega'_{u}}}}(n)$
\end{algorithm}

In plane A, we should choose the optimal routing path from $S$ to
the moving vehicle. We maintain minimum energy consumption of the
h-hop link path from $S$ to node $u$, denoted as ${\Pi_{S-u}}({\text{h}})$,
of which the corresponding minimum cost is ${C_{S-u}}$(h). Firstly,
when hop=1 and for each node $u$ in plane A, we can derive ${C_{S-u}}(1)={\raise0.7ex\hbox{\ensuremath{{{P_{S,u}}\cdot L}}}\!\mathord{\left/{\vphantom{{{P_{S,u}}\cdot L}\rho}}\right.\kern -\nulldelimiterspace}\!\lower0.7ex\hbox{\ensuremath{\rho}}}$,
where $P_{S,u}\leq P_{max}$. Then, when hop is $h\:(h=2,3\ldots,n-1)$,
for each node $u$ and node $v$ in plane A, the minimum energy consumption
is shown as

\begin{equation}
{C_{S-u}}(h)=\min(P_{v,u}\cdot L/\rho+{C_{S-v}}(h-1)).
\end{equation}

And then we can refresh the h-hop path according to 

\begin{equation}
{\Pi_{S-u}}(h)={\Pi_{S-t}}(h-1)+{l_{t,u}},
\end{equation}
where $t=\mathop{\arg\min}\limits _{v\in\varOmega_{A}}(P_{v,u}\cdot L/\rho+{C_{S-v}}(h-1))$.
And we denote the optimal location of the moving vehicle satisfying
(\ref{eq:equal pout}) when node $u$ in plane A communicates with
the moving vehicle, which is the last hop in plane A. So we can have
the minimum energy of node $u$ communicating with moving vehicle,
denoted as $p_{u}^{\textrm{min}}$, which accords with (\ref{eq:equal pout}).
Adding this power to ${C_{S-u}}(n-1)$ of every node $u$ in plane
A, we can choose the minimum energy consumption in plane A with the
n-hop path ${\varPi_{S-u}}(n)$. 

\subsection{\label{subsec:Signal-Propagation-Model-3-1}Routing Algorithm on
Plane B}

\begin{algorithm}[h]
\caption{Dynamic Programming Routing Selection on Plane B}

\ 1:\textbf{ for} \textbf{all} $\varsigma,v\in{\Omega_{B}},{P_{\varsigma,v}}\leqslant{P_{\max}},\varsigma\in\Theta$
\textbf{do}

\ \ \ \ /{*}$\Theta$ is the set of the nodes that is closed to
the trace of \ \ \ \ \ \ 

\ \ \ \ moving vehicle 

\ 2:\ \ \ ${C_{\varsigma-v}}(1)\left|{_{{\Pi_{\varsigma-v}}(1)}}\right.={{P_{\varsigma,v}}\cdot L/\rho}$

\ 3:\textbf{ end for }

\ 4:\textbf{ for} \textbf{all} $u,v,u'\in{\Omega_{B}},{P_{u,v}}\leqslant{P_{\max}},\varsigma\in\Theta$\textbf{do }

\ 5: \ \ \textbf{for} i=2 to m-n \textbf{do} 

\ 6: \ \ \ \ $u'=\mathop{\arg\min}\limits _{u}({C_{\varsigma-u}}(i-1)+{P_{u,v}}\cdot L/\rho)$

\ 7: \ \ \ \ ${C_{\varsigma-v}}(i)={C_{\varsigma-u'}}(i-1)+{P_{u',v}}\cdot L/\rho$

\ 8: \ \ \ \ ${\varPi_{\varsigma-v}}(i)$=${\varPi_{\varsigma-u'}}(i-1)+l_{u',v}$

\ 9: \ \ \textbf{end for }

10:\textbf{ end for }

11:\textbf{ return $\Lambda_{optimal}^{B}(m-n)=\mathop{\arg\min}\limits _{{\Pi_{\varsigma-D}}(n),\varsigma\in\Theta}({C_{\varsigma-D}}(m-n))$}=${\varPi_{\varsigma-v}}(m-n)$
\end{algorithm}

In plane B, as we ignore the energy consumption of the link between
the moving vehicle and the fixed nodes in plane B, there is still
a $(m-n)$-hop path in plane B. We firstly obtain the closest node
set denoted as $\Theta$ to the moving vehicle when the moving vehicle
transmits information to the fixed nodes in plane B. Then we can get
$min(C_{u-D}(m-n)),u\in\Theta$ using a similar algorithm in Section\ref{subsec:Signal-Propagation-Model-3}
to get the minimum energy consumption with the $(m-n)$-hop routing
path ${\Pi_{u-D}}(m-n),u\in\Theta$.

\subsection{\label{subsec:Signal-Propagation-Model-3-1-1}Optimal Routing Path }

\begin{algorithm}[h]
\caption{Find the Optimal Path}

1:\textbf{ for} m=2 to N-1 \textbf{do }

2: \ \ \textbf{for} n=1 to m-1 \textbf{do }

3: \ \ \ \ using Algorithm1 to get the \textbf{$\Lambda_{optimal}^{A}(n)$}

4: \ \ \ \ using Algorithm2 to get the \textbf{$\Lambda_{optimal}^{B}(m-n)$}

5:\textbf{ }\ \ \textbf{end for }

6:\textbf{ end for }

7:\textbf{ return} ${\Lambda_{optimal}}$
\end{algorithm}

We calculate the transmit power of each link according to (\ref{eq:equal pout}),
when $m$ varies from 2 to $N$-1 for each plane with an one-hop path
at least considering the algorithm with the moving vehicle involved.
Then the number of hops in plane A changes from 1 to $m-1$, corresponding
to the number of hops in plane B changes from $m-1$ to 1. Then we
can add up the minimum energy consumption of the entire network. And
the optimal routing path can derived as ${\Lambda_{optimal}}={\varPi_{S-{\omega'_{u}}}}(n)+{\Pi_{u-D}}(m-n)+l_{{\omega'_{u}},u},u\in\Theta,n=1,...,m-1,m=1,...,N-1$.

\subsection{Discussion}

The algorithm described above only considers the optimal routing selection
considering that the moving vehicle must involve information transmission.
This is to say, the moving vehicle satisfies all the possible positions
in order to transmit information. In a practical scenario, the reference
node will take the motion trajectory of the moving vehicle into account.
Moreover, the locations of the source node and the destination node
are also needed to be taken into consideration when deciding whether
or not the moving vehicle should participate in information exchange.

In this paper, because we keep the value of the minimum energy and
the corresponding $m$-hop path selection for each operation, the
computational complexity of the algorithm is $O(N^{4})$ regardless
of the involvement of the moving vehicle. However, for the method
proposed in \cite{Sheikholeslami16energy}, which also considers the
participation of the moving vehicle, the complexity of its algorithm
will be increased to $O(N^{4}logN)$. Therefore, the algorithm complexity
proposed in this paper is lower than that of the MER-EQ algorithm
in \cite{Sheikholeslami16energy} for the scenarios under consideration.

\section{\label{sec:Results-and-Discussions}Results and Discussions}

\begin{figure}
\includegraphics[width=8.5cm]{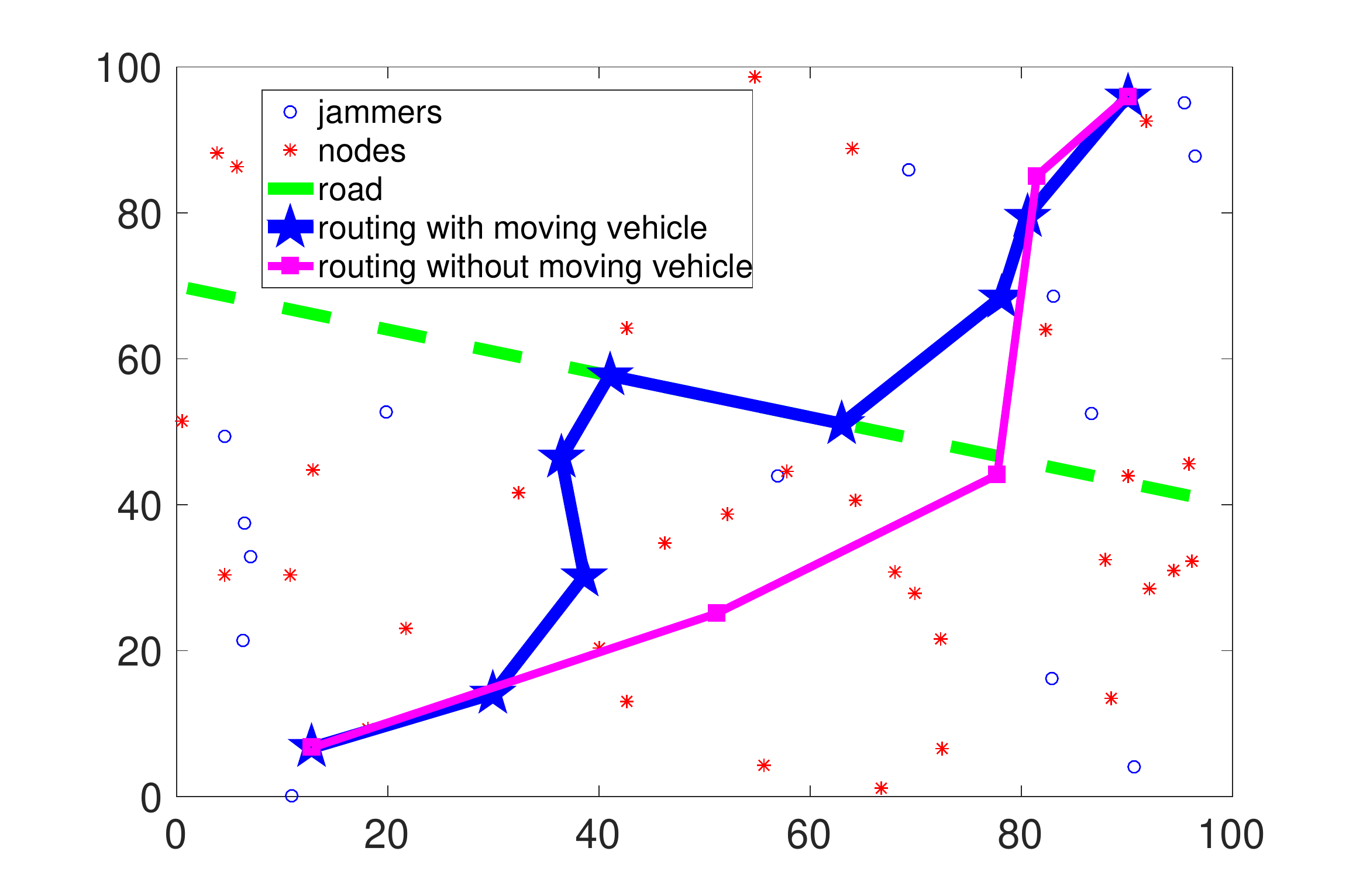}

\caption{\label{fig:alpha=00003D2}Optimal routing path with and without moving
vehicles when $\alpha=2$.}
\end{figure}

Without loss of generality, we assume that the closest system node
to point $(0,0)$ is source $S$, while the closest system node to
point $(100,100)$ is the destination $D$. A snapshot of the network
with an area of $100m\times100m$ is illustrated in Fig. \ref{fig:alpha=00003D2},
where $\lambda_{1}=0.43$, the corresponding number $N=47$, $\lambda_{2}=0.15$,
the corresponding number of jammers is 17, the equation of road is
$3x+10y-700=0$, $P_{\widetilde{u}}=0.1\textrm{W}$, $\alpha=2$,
$P_{\textrm{max}}=15\textrm{W}$ , $T=0.1$, and $L/\rho=1\textrm{s}$
\cite{Ge15Energy}.

In Fig. \ref{fig:alpha=00003D2}, the blue line indicates the selected
optimal path involving moving vehicles, while the pink line is the
selected optimal routing path without the moving vehicles when $\alpha=2$.
Besides, it is found that the minimum energy consumption of the blue
line is about 60\% of the pink one, showing that the routing path
involving the moving vehicles can save much energy compared with the
scenario without the vehicle. What\textquoteright s more, the number
of hops needed in the routing path with moving vehicles is more than
that without moving vehicles, e.g., 8 hops versus 4 hops in Fig. \ref{fig:alpha=00003D2},
indicating that the average energy consumption per node is lower and
thus beneficial in terms of prolonging the service time of the networks.

\begin{figure}
\includegraphics[width=8.5cm]{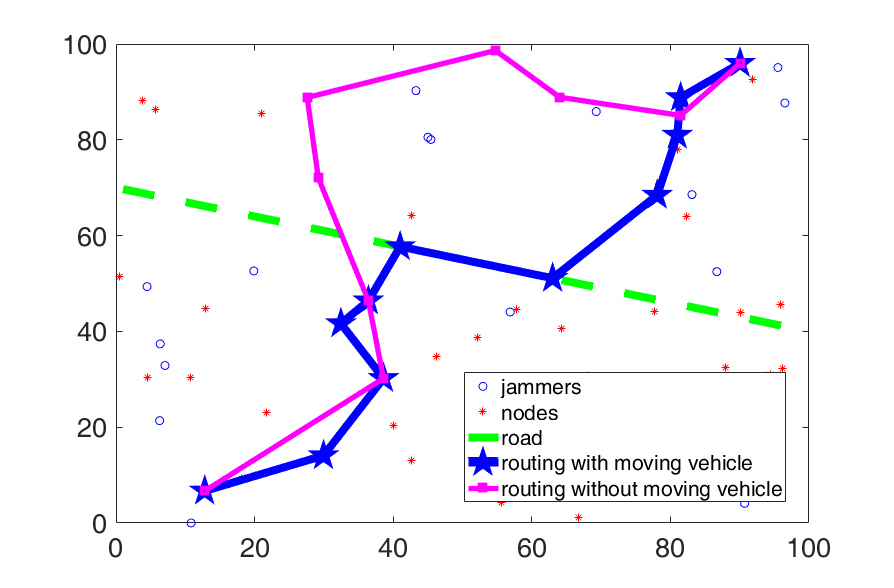}

\caption{\label{fig:alpha=00003D3}Optimal routing path with and without moving
vehicles when $\alpha=3$.}
\end{figure}

\begin{figure}
\includegraphics[width=8.5cm]{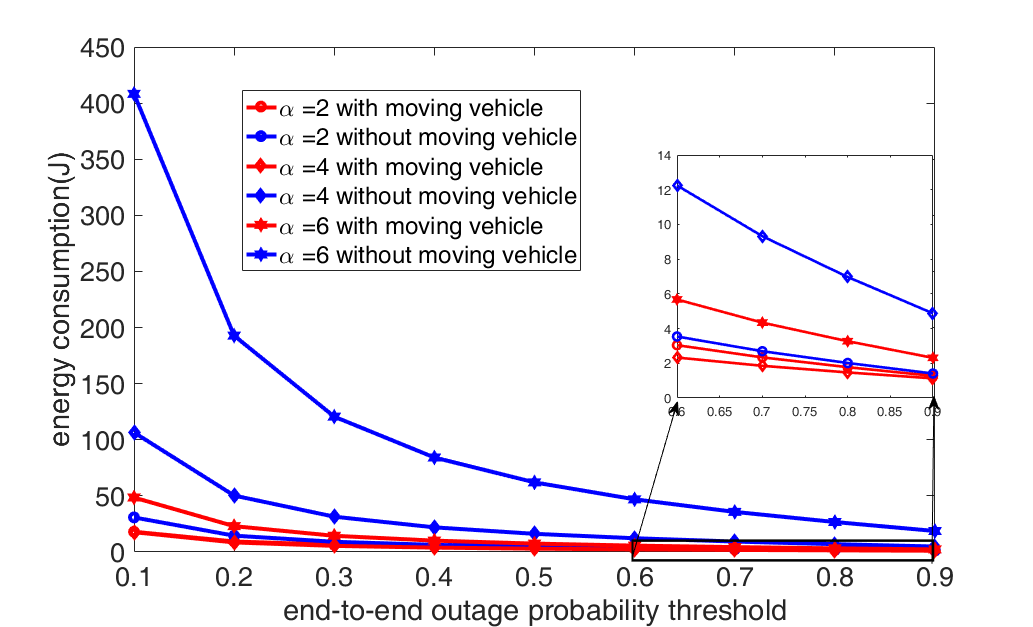}

\caption{\label{fig:energy-vs-treshold}Total energy consumption vs. the end-to-end
outage probability threshold $T$.}
\end{figure}

The optimal routing paths with and without the moving vehicles when
$\alpha=3$ are illustrated in Fig. \ref{fig:alpha=00003D3}. Compared
with Fig. \ref{fig:alpha=00003D2}, the optimal routing path is totally
different. Besides, by utilizing moving vehicles, the total energy
consumption can be saved up to 75\%, which indicates that the path
loss exponent has a great impact on routing path selection and energy
consumption. To further reveal the reason behind, Fig. \ref{fig:energy-vs-treshold}
plots the energy consumption as a function of the end-to-end outage
probability threshold with different path loss exponents.

Without the maximum transmit power constraint, the total energy consumption
versus the end-to-end outage probability threshold $T$ with different
path loss exponents $\alpha$ is depicted in Fig. \ref{fig:energy-vs-treshold}.
It is shown that the energy consumption of the network decreases with
the increase of the end-to-end outage probability threshold, thanks
to a higher requirement of QoS for communications. As for relationship
between the path loss exponents and transmit power of each link, we
can obtain $p_{u,v}^{\textrm{out}}\approx1-\exp\left(-\frac{{d_{u,v}}^{\alpha}\cdot\gamma}{P_{u,v}}\cdot\sum\limits _{\widetilde{u}}{{P_{\widetilde{u}}}d_{\widetilde{u},v}^{-\alpha}}\right)$
from the fact that $e^{x}\geq1+x$ for $x\geq0$. And then we can
obtain $P_{u,v}(\alpha)\propto{d_{u,v}}^{\alpha}\cdot\gamma\cdot\sum\limits _{\widetilde{u}}{{P_{\widetilde{u}}}d_{\widetilde{u},v}^{-\alpha}}=\gamma\cdot{\sum\limits _{\widetilde{u}}{{P_{\widetilde{u}}}(\frac{{d_{u,v}}}{{d_{\widetilde{u},v}}})}^{\alpha}}$
based on \eqref{eq:equal pout}. $\frac{{d_{u,v}}}{{d_{\widetilde{u},v}}}$
has a different effect on the transmit power $P_{u,v}$. For instance,
when $\frac{{d_{u,v}}}{{d_{\widetilde{u},v}}}>1$ , the transmit power
increases with the path loss exponents, and decreases the other way
round. Thus, we can find that the sum energy consumption when $P_{u,v}$
is higher than when $\alpha=4$, but lower than when $\alpha=6$.
The same can be concluded from Figs. \ref{fig:alpha=00003D2} and
\ref{fig:alpha=00003D3}. The minimum energy consumption involving
the moving vehicle in Fig. \ref{fig:alpha=00003D3} is lower than
that in Fig. \ref{fig:alpha=00003D2}.

\begin{figure}
\includegraphics[width=8.5cm]{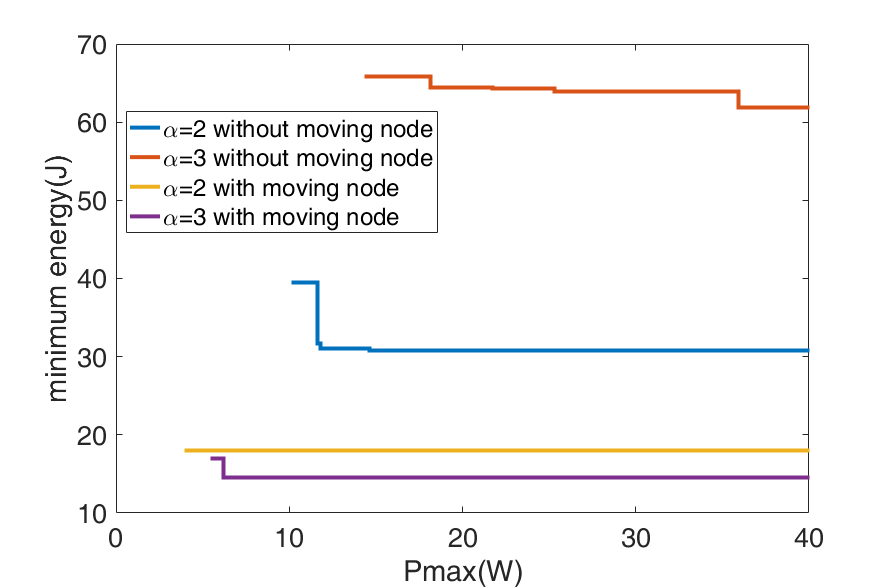}

\caption{\label{fig:Pmax-vs-energy}Minimum energy consumption vs. the maximum
power constrain.}
\end{figure}

Fig. \ref{fig:Pmax-vs-energy} shows the minimum network energy consumption
as a function of the maximum power constrain $P_{\max}$ with different
path loss exponents when $T=0.1$. It is found that the minimum network
energy consumption decreases with the increase of $P_{\max}$, indicating
that a strict QoS constraint, i.e., the configuration of $T$, makes
it more difficult to transmit information in a small number of hops,
and thus the system requires a greater number of hops when $P_{\max}$
is low. Moreover, when $P_{\max}$ exceeds a certain value, the minimum
network energy consumption remains constant. By contrast, there is
no proper routing path between $S$ and $D$, when $P_{\max}$ is
lower than a given value denoted by $\overline{P_{\max}}$. It is
also noted that the value of $\overline{P_{\max}}$ is smaller when
transferring information with the moving vehicles than without the
moving vehicle.

\section{\label{sec:Results-and-Discussions-1}Conclusions and Future Work}

In this paper, we investigated the optimal routing path design in
suburban areas by jointly considering the per-node maximum transmit
power constraint, QoS and energy-efficient communications. In our
model, moving vehicles are used to assist in information transportation.
A three-step algorithm was proposed to find the optimal routing path
with a computational complexity of $O(N^{4})$. Besides, results were
presented to show that with the assistance of a moving vehicle, the
total energy consumed can be reduced greatly. We also studied the
impact on routing path design and energy consumed caused by the path
loss exponent, maximum transmit power constrain and QoS requirement.
In our future work, a multi-point-topoint transmission method will
be considered.

\bibliographystyle{IEEEtran}
\bibliography{hmd,hmd2}

\begin{thebibliography}{10}
\providecommand{\url}[1]{#1}
\csname url@samestyle\endcsname
\providecommand{\newblock}{\relax}
\providecommand{\bibinfo}[2]{#2}
\providecommand{\BIBentrySTDinterwordspacing}{\spaceskip=0pt\relax}
\providecommand{\BIBentryALTinterwordstretchfactor}{4}
\providecommand{\BIBentryALTinterwordspacing}{\spaceskip=\fontdimen2\font plus
\BIBentryALTinterwordstretchfactor\fontdimen3\font minus
  \fontdimen4\font\relax}
\providecommand{\BIBforeignlanguage}[2]{{%
\expandafter\ifx\csname l@#1\endcsname\relax
\typeout{** WARNING: IEEEtran.bst: No hyphenation pattern has been}%
\typeout{** loaded for the language `#1'. Using the pattern for}%
\typeout{** the default language instead.}%
\else
\language=\csname l@#1\endcsname
\fi
#2}}
\providecommand{\BIBdecl}{\relax}
\BIBdecl

\bibitem{Ericsson15White}
Ericsson, \emph{Ericsson White Paper}, January 2016, available at:
  https://www.ericsson.com/assets/local/publications/white-papers/wp\_iot.pdf.

\bibitem{wei2016}
W.~Xiang, K.~Zheng, and X.~Shen, \emph{5G Mobile Communications}.\hskip 1em
  plus 0.5em minus 0.4em\relax Springer, 2016.

\bibitem{Ericsson15Mobility}
Ericsson, \emph{Ericsson Mobility Report}, November 2015, available at:
  http://www.ericsson.com/res/docs/2015/mobility-report/ericsson-mobility-report-nov-2015.pdf.

\bibitem{Rahama16A}
M.~T. Rahama, M.~Hossen, and M.~M. Rahman, ``A routing protocol for improving
  energy efficiency in wireless sensor networks,'' in \emph{Proc. ICEEICT}, pp.
  1--6.

\bibitem{Huynh16Delay}
T.~T. Huynh, A.~V. Dinh-Duc, and C.~H. Tran, ``Delay-constrained
  energy-efficient cluster-based multi-hop routing in wireless sensor
  networks,'' \emph{Journal of Communications and Networks}, vol.~18, no.~4,
  pp. 580--588, Sep. 2016.

\bibitem{Tran12Minimum}
D.~H. Tran and D.~S. Kim, ``Minimum latency and energy efficiency routing with
  lossy link awareness in wireless sensor networks,'' in \emph{2012 9th IEEE
  International Workshop on Factory Communication Systems}, pp. 75--78.

\bibitem{Jan17A}
N.~Jan, N.~Javaid, Q.~Javaid, N.~A. Alrajeh, M.~Alam, Z.~A. Khan, and I.~A.
  Niaz, ``A balanced energy consuming and hole alleviating algorithm for
  wireless sensor networks,'' \emph{IEEE Access}, vol.~PP, no.~99, pp. 1--1,
  2017.

\bibitem{Bazzi13Vehicle}
A.~Bazzi, B.~M. Masini, A.~Zanella, and G.~Pasolini, ``Vehicle-to-vehicle and
  vehicle-to-roadside multi-hop communications for vehicular sensor networks:
  Simulations and field trial,'' in \emph{Proc. ICC}, 2013, pp. 515--520.

\bibitem{Korkmaz06An}
G.~Korkmaz, E.~Ekici, and F.~Ozguner, ``An efficient fully ad-hoc multi-hop
  broadcast protocol for inter-vehicular communication systems,'' in
  \emph{Proc. ICC}, vol.~1, 2006, pp. 423--428.

\bibitem{Tonguz07Broadcasting}
O.~Tonguz, N.~Wisitpongphan, F.~Bait, P.~Mudaliget, and V.~Sadekart,
  ``Broadcasting in vanet,'' in \emph{2007 Mobile Networking for Vehicular
  Environments}, pp. 7--12.

\bibitem{Ge16Vehicular}
X.~Ge, H.~Cheng, G.~Mao, Y.~Yang, and S.~Tu, ``Vehicular communications for 5g
  cooperative small-cell networks,'' \emph{IEEE Transactions on Vehicular
  Technology}, vol.~65, no.~10, pp. 7882--7894, Apr., 2016.

\bibitem{Sheikholeslami14Jamming}
A.~Sheikholeslami, M.~Ghaderi, H.~Pishro-Nik, and D.~Goeckel, ``Jamming-aware
  minimum energy routing in wireless networks,'' in \emph{Proc. ICC}, 2014, pp.
  2313--2318.

\bibitem{Zhao16An}
M.~Zhao, I.~W.~H. Ho, and P.~H.~J. Chong, ``An energy-efficient region-based
  rpl routing protocol for low-power and lossy networks,'' \emph{IEEE Internet
  of Things Journal}, vol.~3, no.~6, pp. 1319--1333, Jul., 2016.

\bibitem{Wang96Quality}
W.~Zheng and J.~Crowcroft, ``Quality-of-service routing for supporting
  multimedia applications,'' \emph{IEEE Journal on Selected Areas in
  Communications}, vol.~14, no.~7, pp. 1228--1234, Aug. 1996.

\bibitem{Sheikholeslami16energy}
A.~Sheikholeslami, M.~Ghaderi, H.~Pishro-Nik, and D.~Goeckel,
  ``Energy-efficient routing in wireless networks in the presence of jamming,''
  \emph{IEEE Transactions on Wireless Communications}, vol.~15, no.~10, pp.
  6828--6842, Oct. 2016.

\bibitem{Ge15Energy}
X.~Ge, S.~Tu, T.~Han, Q.~Li, and G.~Mao, ``Energy efficiency of small cell
  backhaul networks based on gauss-markov mobile models,'' \emph{IET Networks},
  vol.~4, no.~2, pp. 158--167, Mar., 2015.

\end{thebibliography}

\end{document}